\documentclass[12pt]{article}

\usepackage{amsmath, setspace}
\usepackage{amssymb, xspace}

\usepackage{graphicx, rotating}

\usepackage{color} 
\usepackage{colortbl}

\usepackage{overpic}
\usepackage{times}
\usepackage[scaled]{helvet}
\usepackage{array}
\usepackage{caption}
\usepackage{rotating}

\usepackage{hyperref}
\usepackage{times}
\usepackage{cite}
\usepackage[table,rgb]{xcolor}

\usepackage[misc,geometry]{ifsym} 

\usepackage[labelfont=bf,labelsep=period,justification=raggedright]{caption}

\makeatletter
\renewcommand{\@biblabel}[1]{\quad#1.}
\makeatother

\date{}

\newcommand{\supp}{Methods\xspace} 

\newcommand{\fg}{\textcolor{citecolor}{Fig} \ref}
\newcolumntype{P}[1]{>{\centering\arraybackslash}p{#1}}

\definecolor{citecolor}{rgb}{0.05, 0.05, 0.65}
\definecolor{linkcolor}{rgb}{0.2, 0.05, 0.05}
\definecolor{lightGray}{gray}{0.9}
\hypersetup{
colorlinks=true, 
citecolor=citecolor,
linkcolor=citecolor,
urlcolor=citecolor
}

\newcommand{\pana}{({\bf a})\xspace}
\newcommand{\panb}{({\bf b})\xspace}
\newcommand{\panc}{({\bf c})\xspace}
\newcommand{\pand}{({\bf d})\xspace}
\newcommand{\pane}{({\bf e})\xspace}

\newcommand{\figs}[1]{{Fig S#1}}

\newcommand{\blue}{\textcolor{black}}

\newcommand{\rp}{R\textsubscript{P}\xspace	} 
\newcommand{\rt}{R\textsubscript{T}\xspace	}

\begin{document}


\def \simParadox { {\bf The fraction of total protein variance explained
     by scaled mRNA levels is not informative about the across-tissues
     variance explained by scaled mRNA levels.} 
     \blue{ \pana mRNA levels 
     	correlate with measured protein levels ($R_T = 0.33$ over all measured mRNAs
       and proteins across 12 different tissues).  
    \panb Protein levels versus mRNA levels scaled by the median protein-to-mRNA ratio (PTR); 
    the only change from  panel (a) is the scaling of mRNAs which considerably 
    improves the correlation.}   
  \panc A subset of 100 genes are
   used to illustrate an example Simpson's paradox: regression lines
   reflect within-gene and across-tissues variability.
   Despite the fact that the overall correlation between scaled mRNA
   and measured protein levels is large and positive $R_T = 0.89$, for
   any single gene in this set, mRNA levels scaled by the median PTR ratio are \emph{not}
   correlated to the corresponding measured protein levels ($R_P \approx 0$).
   \blue{\pand Cumulative distributions of across-tissues
	scaled mRNA-protein correlations (\rp) for 3 datasets
	 \citep{wilhelm14,kim2014draft,edfors2016gene}. The smooth curves correspond to all
	 quantified proteins by shotgun proteomics while the dashed curves correspond to
	  a subset of proteins quantified in a small targeted dataset\cite{edfors2016gene}. 
	  The vertical lines show the corresponding overall (conflated) correlation between 
	  scaled mRNA levels and protein levels, \rt.}    
    See \supp and \figs{1}. 
}

\def \Reliability { {\bf Data reliability crucially influences estimates of  transcriptional and post-transcriptional regulation across-tissues.  }
   \pana The within-study reliability -- defined as the fraction of the  measured variance due to the signal -- of relative mRNA levels is estimated as the correlation between the mRNA levels measured in the twelve different tissues. Estimates for the levels of each transcript measured in different subjects were correlated (averaging across the 12 tissue-types) and the results for all analyzed transcripts displayed as a distribution for each RNA dataset \citep{ENCODE2012landscape, fagerberg2014analysis}. 
   \panb The  within-study reliability of relative protein levels is estimated as the correlation between the protein levels measured in 12 different tissues \citet{wilhelm14,kim2014draft}. Within each dataset, separate estimates for each protein were derived from non-overlapping sets of peptides and were correlated (averaging across the 12 tissue-types) and the results for all analyzed proteins displayed as a distribution; see Methods. 
    \panc The  across-study reliability of mRNA was estimated by correlating estimates as in (a) but these estimates came from different studies \citep{ENCODE2012landscape} and    \citep{fagerberg2014analysis}.
    \pand The across-study reliability of proteins was estimated by correlating estimates as in (b) but these estimates came from different studies \citep{wilhelm14} and \citep{kim2014draft}.
    \pane The fraction of across-tissues protein variance that can be explained by mRNA levels is plotted as a function of the reliability of the estimates of mRNA and protein levels, given an empirical mRNA/protein correlation of 0.29.  The red Xs correspond to two estimates of reliability of the mRNA and protein measurements computed from both independent mRNA and protein datasets.  }

\def \rPTR { {\bf Concerted variability in the relative protein-to-RNA (rPTR) ratio of functional gene-sets across tissue-types} 
      \pana mRNAs coding for the ribosomal proteins, NADH dehydrogenase and respiratory proteins have higher protein-to-mRNA ratios in kidney as compared to the median across the other 11 tissues (FDR $< 1\%$). In contrast mRNAs genes functioning in Rac GTPase binding have lower protein-to-mRNA ratios (FDR $< 1\%$).  
    \panb The stomach also shows significant rPTR variation, with low rPTR for the ribosomal proteins and high rPTR for tRNA-aminoacylation (FDR $< 1\%$).  
    \panc Summary of rPTR
    variability, as depicted in panel (a-b), across all tissues and
    many gene ontology (GO) terms. Metabolic pathways and functional
    gene-sets that show statistically significant (FDR $< 1\%$)
    variability in the relative protein-to-mRNA ratios across the 12 tissue
    types. All data are displayed on a $\log_{10}$ scale, and functionally related gene-sets are marked with the same color. 
     \pand The reproducibility of rPTR estimates across estimates from different studies is estimated as the correlation between the median rPTRs for GO terms showing significant enrichment as shown in panels (a-c).  
     See \supp, \figs{2} and \figs{3}.
}

\def \consensus { {\bf Deriving a consensus protein dataset for improved quantification of human tissue proteomes  } 
    We compiled a consensus protein dataset by merging data from \citet{wilhelm14} and \citet{kim2014draft} as described in Methods. The relative protein levels estimated from \citep{wilhelm14}, \citep{kim2014draft}, and the consensus dataset were correlated to mRNA levels from \citet{fagerberg2014analysis} \pana or to   mRNA levels from \citep{ENCODE2012landscape} \panb.  The correlations are shown as a function of the median correlation between protein estimates from \citet{wilhelm14} and \citet{kim2014draft}. The consensus dataset exhibits the highest correlations, suggesting that it has averaged out some of the noise in each dataset and provides a more reliable quantification of of human tissue proteomes.
    \blue{\panc  The datasets from \citet{wilhelm14}, from \citet{kim2014draft}, and the consensus dataset were evaluated by comparison to a targeted MS validation dataset quantifying 33 proteins over 5 tissues \citep{edfors2016gene}. The similarity for each dataset was quantified by the mean squared error (MSE) relative to the targeted MS validation data using 124 protein/tissue measurements that were observed in all datasets.  The MSEs are reported for each of the five tissues and for all 5 tissues combined; they indicate that the consensus data have the best agreement with the validation dataset.}   
}

\def  \covarage_Table { {\bf Proteome coverage.}  Number of quantified proteins and peptides in 3 datasets: \citet{wilhelm14}, \citet{kim2014draft}, and the consensus dataset derived here. The consensus data have the highest coverage and the best agreement with the validation dataset, \fg{consensus}. 
}

\def  \Simpson_Supp { {\bf The total protein variance explained by scaled mRNA levels is not indicative of the correlations between mRNA and protein fold-changes across the corresponding tissue pairs.} 
(a-c, top row), protein versus mRNA in kidney, liver and prostate. 
(d-f, middle row) protein versus scaled mRNA in kidney, liver and prostate. The only difference from the top row is that the mRNA was scaled by the median PTR.  
(g-i, bottom row) protein fold changes versus the corresponding mRNA fold changes between the tissues indicated on the top. While scaled mRNA is predictive of the absolute protein levels the accuracy of these predictions does not generally reflect the accuracy of protein fold-changes across tissues that are predicted from the corresponding mRNA fold-changes. RNA fold changes in (g-i, bottom row) were computed between the mRNA levels without PTR scaling.  
}

\def  \GOterm_Corrs  { {\bf Fraction of across-tissues variability in protein levels explained by RNA variability for different functional gene sets.} 
\pana The distributions of across-tissues correlations for gene sets defined by the gene ontology are shown as boxplots. The reliability of RNA and protein are estimated as the correlations between estimates from different datasets.     
\panb For each gene set, the median RNA-protein correlation was corrected by the median reliabilities and the results shown as a boxplot.  
Differences between RNA-protein correlations for different gene-sets cannot be explained simply by differences in the reliabilities. 
}

\def \rptrCorrs {{\bf Reproducibility of rPTR ratios estimated from different datasets} The x-axes shows estimates from Wilhelm et al. \citep{wilhelm14} and the y-axes estimates from Kim et al. \citep{kim2014draft}.
}

\def \rPTRrGOTable { {\bf Estimates of relative protein-to-RNA (rPTR) ratio for GO terms reproduce across different datasets} Pearson correlations between two estimates of the median rPTR ratios for all GO terms indicate reproducible effects in all tissues. As in \fg{reliability}, rPTR estimates are derived using independent data sources.  The lower and upper estimates are the endpoints of the 95\% confidence interval.  }

\def  \rPTRrGenesTable { {\bf Estimates of relative protein-to-RNA (rPTR) ratio for genes reproduce across different datasets}  Correlations between the two estimates of rPTR ratios for all genes indicate reproducible effects in all tissues. The rPTR ratios were estimated independently from different datasets (as in \fg{reliability}).  The lower and upper estimates are the endpoints of the 95\% confidence interval. 
}

\def  \mseTab { {\bf Data quality and coverage.}  The datasets from \citet{wilhelm14}, from \citet{kim2014draft}, and the consensus dataset were evaluated by comparison to a targeted MS validation dataset \citep{edfors2016gene}. The similarity for each dataset was quantified by the mean squared error (MSE) and median absolute deviation (MAD) from the targeted MS validation data using 124 protein/tissue measurements that were observed in all data sets.  We also report the coverage of each data set as the total number of observed protein/tissue values. The consensus data have the highest coverage and the best agreement with the validation dataset. 
}

\let\citep=\cite
\let\citet=\cite

\title{Post-transcriptional regulation across human tissues}

\author{ 
Alexander Franks$^{1,2}$,
Edoardo Airoldi$^{3,4}$,
Nikolai Slavov$^{5,6,}$\textsuperscript{ \Letter} 
\\  \\ 
\normalsize{$^{1}$Department of Statistics, University of Washington, Seattle, WA 98195, USA}\\
\normalsize{$^{2}$Current address: Department of Statistics, University of California, Santa Barbara, Santa Barbara, CA}\\
\normalsize{$^{3}$Department of Statistics, Harvard University, Cambridge, MA 02138, USA} \\
\normalsize{$^{4}$Broad Institute of MIT and Harvard University, Cambridge, MA 02142, USA} \\
\normalsize{$^{5}$Department of Bioengineering, Northeastern University, Boston, MA 02115, USA}\\
\normalsize{$^{6}$Department of Biology, Northeastern University, Boston, MA 02115, USA} \\[1em]
 \Letter \; Correspondence: \href{mailto:nslavov@alum.mit.edu}{nslavov@alum.mit.edu}\\ 
}

\maketitle
\vspace{2cm}

\section*{Abstract}
Transcriptional and post-transcriptional regulation shape tissue-type-specific proteomes, but their relative contributions remain contested. 
Estimates of the factors determining protein levels in human tissues do not distinguish between
$(i)$   the factors determining the variability between the abundances of different proteins, i.e., mean-level-variability and,
$(ii)$  the factors determining the physiological variability of the same protein across different tissue types, i.e., across-tissues variability. We sought to estimate the contribution of transcript levels to these two orthogonal sources of variability, and found that scaled mRNA levels can account for most of the mean-level-variability but not necessarily for across-tissues variability. The reliable quantification of the latter estimate is limited by substantial measurement noise. However,  protein-to-mRNA ratios exhibit substantial across-tissues variability that is functionally concerted and reproducible across different datasets, suggesting extensive post-transcriptional regulation. These results caution against estimating protein fold-changes from mRNA fold-changes between different cell-types, and highlight the contribution of post-transcriptional regulation to shaping tissue-type-specific proteomes. 
\bigskip
\begin{spacing}{1.4}

\section*{Author Summary}
The identity of human tissues depends on their protein levels. Are tissue protein levels set largely by corresponding mRNA levels or by other (post-transcriptional) regulatory mechanisms? We revisit this question based on statistical analysis of mRNA and protein levels measured across human tissues. We find that for any one gene, its protein levels across tissues are poorly predicted by its mRNA levels, suggesting tissue-specific post-transcriptional regulation. In contrast, the overall protein levels are well predicted by scaled mRNA levels. We show how these speciously contradictory findings are consistent with each other and represent the two sides of Simpson's paradox.

\section*{Introduction}
The relative ease of measuring mRNA levels has facilitated numerous investigations of how cells regulate their gene expression across different pathological and physiological conditions \citep{sorlie2001gene, Slavov_2009, spellman1998comprehensive, Slavov_batch_ymc, Slavov_emc, djebali2012landscape}. However, often the relevant biological processes depend on protein levels, and mRNA levels are merely proxies for protein levels \citep{Alberts_2014}. If a gene is regulated mostly transcriptionally, its mRNA level is a good proxy for its protein level. Conversely,  post-transcriptional regulation can set 
protein levels independently from mRNA levels, as in the cases of classical regulators of development \citep{kuersten2003power}, cell division \citep{Reed_SI_1996translational, polymenis1997coupling} and metabolism \citep{daran2007fluxes, Slavov_exp}. Thus understanding the relative contributions of transcriptional and post-transcriptional regulation is essential for understanding their trade-offs and the principles of biological regulation, as well as for assessing the feasibility of using mRNA levels as proxies for protein levels.

Previous studies have considered single cell-types and conditions in studying variation in absolute mRNA and protein levels genome-wide, often employing unicellular model organisms or mammalian cell cultures \citep{gygi1999correlation, smits2014global, schwanhausser2011global, li2014system, Csardi2015, jovanovic2015dynamic, cheng2016differential}. 
 However, analyzing per-gene variation in relative mRNA and protein expression across different tissue types in a multicellular organism presents a potentially different and critical problem which cannot be properly addressed by examining only genome-scale correlations between mRNA and protein levels. \citet{wilhelm14,kim2014draft,edfors2016gene}  have measured protein levels across human tissues, thus providing valuable datasets for analyzing the regulatory layers shaping tissue-type-specific proteomes. The absolute levels of proteins and mRNAs in these datasets correlate well, highlighting that highly abundant proteins have highly abundant mRNAs. Such correlations between the absolute levels of mRNA and protein mix/conflate many sources of variation, including variability between the levels of different proteins, variability within the same protein across different conditions and cell-types, and the variability due to measurement error and technological bias.

However, these different sources of variability have very different biological interpretations and implications. A major source of variability in protein and mRNA data arises from differences between the levels of mRNAs and proteins corresponding to different genes. That is, the mean levels (averaged across tissue-types) of different proteins and mRNAs vary widely.  We refer to this source of variability as \emph{mean-level variability}. This mean-level variability reflects the fact that some proteins, such as ribosomal proteins, are highly abundant across all profiled tissues while other proteins, such as cell cycle and signaling regulators, are orders of magnitude less abundant across all profiled conditions \citep{wilhelm14}.  Another principal source of variability in protein levels, intuitively orthogonal to the mean-level variability, is the variability within a protein across different cell-types or physiological conditions and we refer to it as \emph{across-tissues variability}. The  across-tissues  variability is usually much smaller in magnitude, but may be the most relevant source of variability for understanding different phenotypes across cells-types and physiological conditions.

Here, we sought to separately quantify the contributions of transcriptional and post-transcriptional regulation to the mean-level variability and to the across-tissues  variability across human tissues. Our results show that the much of the mean-level protein variability can be explained well by mRNA levels while across-tissues protein variability is poorly explained by mRNA levels; much of the unexplained variance is due to measurement noise but some of it is reproducible across datasets and thus likely reflects post-transcriptional regulation.  These results add to previous results in the literature \citep{gygi1999correlation, schwanhausser2011global, li2014system, wilhelm14,  jovanovic2015dynamic, Csardi2015, smits2014global, edfors2016gene} and suggest that the post-transcriptional regulation is a significant contributor to shaping tissue-type specific proteomes in human. 

\section*{Results}
\subsection*{The correlation between absolute mRNA and protein levels conflates distinct sources of variability} 
We start by outlining the statistical concepts underpinning the common correlational analysis and depiction  \citep{gygi1999correlation, schwanhausser2011global, wilhelm14, Csardi2015} of estimated absolute protein and mRNA levels  as displayed in \fg{figure1}a,b. The correlation between the absolute mRNA and protein levels of different genes and across different tissue-types has been used to estimate the level at which the protein levels are regulated \citep{wilhelm14,edfors2016gene}.

One measure reflecting the post-transcriptional regulation of a gene is its protein to mRNA ratio, which is sometimes referred to as a gene's ``translational efficiency''.  Since this ratio also reflects other layers of regulation, such as protein degradation and noise \citep{jovanovic2015dynamic}, we will refer to it descriptively as \emph{protein-to-mRNA} (PTR) ratio. If the across-tissues  variability of a gene is dominated by transcriptional regulation, its PTR in different tissue-types will be a gene-specific constant. Based on this idea, \citet{wilhelm14,edfors2016gene} estimated these protein-to-mRNA ratios and suggested that the median PTR for each gene can be used to scale its tissue-specific mRNA levels and that this ``scaled mRNA'' predicts accurately tissue-specific protein levels.

Indeed, mRNA levels scaled by the corresponding median PTR explain large fraction of the total protein variance ($R_T^2=0.70$, across all measured proteins, \fg{figure1}a,b) as previously observed \citep{schwanhausser2011global, wilhelm14,edfors2016gene}. However, this high $R_T^2$ does not indicate concordance for across-tissues variability of mRNAs and proteins.  $R_T^2$ quantifies the fraction of the total protein variance explained by mRNA levels between genes and across tissue-types; thus, it conflates the mean-level variability with the across-tissues  variability. This conflation is shown schematically in \fg{figure1}c for a subset of 100 genes measured across 12 tissues.  The across-tissues  variability is captured by the variability \emph{within} the regression fits while the mean-level variability is captured by the variability \emph{between} the regression fits. 
 
Such aggregation of distinct sources of variability, where different subgroups of the data show different trends, may lead to counter-intuitive results and incorrect conclusions, and is known as the Simpson's or amalgamation paradox \citep{blyth1972simpson}. 
To illustrate the Simpson's paradox in this context, we depicted a subset of genes for which the measured mRNA and protein levels are unrelated across-tissues while the mean-level variability still spans the full dynamic range of the data, \fg{figure1}c.  For this subset of genes, the overall (conflated/amalgamated) correlation is large and positive, despite the fact that all across-tissues (within-gene) trends are close to zero.  This counter-intuitive result is possible because the conflated correlation is dominated by the variability with larger dynamical range, in this case the mean-level variability.  This conceptual example using data from \citet{wilhelm14} demonstrates that $R^2_T$ is not necessarily informative about the across-tissues  variability, i.e., 
 the protein variance explained by scaled mRNA \emph{within} a gene ($R^2_P$). Thus the conflated correlation is not generally informative about the level --- transcriptional or post-transcriptional --- at which across-tissues variability is regulated. This point is also illustrated in \figs{1} with data for all quantified genes: The correlations between scaled mRNA and measured protein levels are not informative for the correlations between the corresponding relative changes in protein and mRNA levels.   
 
\blue{To further illustrate this point with more datasets, \fg{figure1}d displays the cumulative distributions of across-tissues mRNA-protein correlations (\rp) for all proteins quantified across the large shotgun datasets \citep{wilhelm14, kim2014draft}, as well as the corresponding conflated correlations between scaled mRNA and protein levels (\rt). This depiction demonstrates that \rt are not representative for \rp. To extended this analysis to protein levels measured by targeted MS \citep{edfors2016gene}, we plotted the distributions of across-tissues mRNA-protein correlations (\rp) for the subset of 33 genes quantified across all datasets \citep{wilhelm14, kim2014draft, edfors2016gene}; see dotted curves in \fg{figure1}d. These genes were selected to have larger variance across tissues  \citep{edfors2016gene} and have relatively higher across-tissues correlations, especially in the data by \citet{kim2014draft, edfors2016gene}. Nevertheless, all datasets include low and even negative across-tissues correlations (\rp) and very high conflated correlations between scaled mRNA and protein levels (\rt),  \fg{figure1}d. These results underscore the weak connection between \rp and \rt even for a carefully selected and measured subset of mRNAs and proteins.}

The across-tissues variability has a dynamic range of about $2-10$ fold and is thus dwarfed by the $10^3-10^4$ fold dynamic range of abundances across different proteins. While across-tissues variability is smaller than mean-level variability, it is exactly the across-tissues variability that contributes to the biological identity of each tissue type and we focus the rest of our analysis to factors regulating the across-tissues protein variability.

\subsection*{Estimates of  transcriptional and post-transcriptional regulation across-tissues depend strongly on data reliability}
 Next, we sought to  estimate the fractions of  across-tissues  protein variability due to transcriptional regulation and to post-transcriptional regulation. This estimate depends crucially on noise in the mRNA and protein data, from sample collection to measurement error.  
Both RNA-seq \citep{marioni2008rna, seqc2014comprehensive} and mass-spectrometry \citep{schwanhausser2011global, albert-heck2012protease} have relatively large and systematic error in estimating \emph{absolute} levels of mRNAs and proteins, i.e., the ratios between different proteins/mRNAs. These errors originate from DNA sequencing GC-biases, and variations in protein digestion and peptide ionization. However, relative quantification of the same gene across tissue-types by both methods can be much more accurate since systematic biases are minimized when taking ratios between the intensities/counts of the same peptide/DNA-sequence measured in different tissue types \citep{ong2002stable, blagoev2004temporal, seqc2014comprehensive, jovanovic2015dynamic}. It is this relative quantification that is used in estimating across-tissues  variability, and we start by estimating the reliability of the relative quantification across human tissues, \fg{reliability}a-d. Reliability is defined as the fraction of the observed/empirical variance due to signal. Thus reliability is proportional to the signal strength and decreases with the noise levels. 

\blue{To estimate the \emph{within study} reliability of mRNA levels, we took advantage that each mRNA dataset contains data from multiple subjects. We split the subjects in each dataset into two subsets, each of which containing measurements for all 12 tissues from several subjects. The levels of each mRNA were estimated from each subset by averaging across subjects and the estimates from the two subsets correlated, \fg{reliability}a. These correlations provide estimates for the reliability of each mRNA and their median provides a global estimate for the reliability of relative RNA measurement, not taking into account noise due to sample collection and processing. }

To estimate the \emph{within study} reliability of protein levels, we computed separate estimates of the relative protein levels within a dataset. For each protein, Estimate 1 was derived from 50 \% of the quantified peptides and Estimate 2 from the other 50 \%.  
 Since much of the analytical noise related to protein digestion, chromatographic mobility and peptide ionization is peptide-specific, such non-overlapping sets of of peptides provide mostly, albeit not completely,  independent estimates for the relative protein levels. The correlations between the estimates for each protein (averaging across 12 tissues) are displayed as a distribution in \fg{reliability}b.

In addition to the \emph{within study} measurement error, protein and mRNA estimates can be affected by study-dependable variables such as sample collection and data processing. To account for these factors, we estimated \emph{across study} reliability by comparing estimates for relative protein and mRNA levels derived from independent studies, \fg{reliability}c-d. For each gene, we estimate the reliability for each protein by computing the the empirical correlation between mRNA abundance reported by the ENCODE \citep{ENCODE2012landscape} and by   \citep{fagerberg2014analysis}.  
The correlations in \fg{reliability}c have much broader distribution than the within-study correlations, indicating that much of the noise in mRNA estimates is study-dependent.

To estimate the  \emph{across study} reliability of protein levels, we compared the protein levels estimated from data published by \citet{wilhelm14} and \citet{kim2014draft}. To quantify protein abundances, \citet{wilhelm14} used iBAQ scores and \citet{kim2014draft} used spectral counts. To ensure uniform processing of the two datasets, we downloaded the raw data and analyzed them with maxquant using identical settings, and estimated protein abundances in each dataset using iBAQ; see Methods. The corresponding  estimates for each protein were correlated to estimate their reliability. 
Again, the correlations depicted in \fg{reliability}d have a much broader distribution compared to the within-study protein correlations in \fg{reliability}b, indicating that, as with mRNA, the vast majority of the noise is study-dependent.  As a representative estimate of the reliability of protein levels, we use the median of the across tissue correlations from Figures \ref{reliability}c-d.

The across tissues correlations and the reliability of the measurements can be used to estimate the across tissues variability in protein levels that can be explained by mRNA levels (i.e., transcriptional regulation) as shown in \fg{reliability}e; see Methods. As the reliabilities of the protein and the mRNA estimates decrease, the noise sensitivity of the estimated transcriptional contribution increases. 
Although the average across-tissues mRNA protein correlation was only $0.29$ ($R^2=0.08$), the data are consistent with approximately $50\%$ of the variance being explained by transcriptional regulation and approximately $50\%$ coming from post-transcriptional regulation; see \figs{2} for reliability-corrected estimates for specific functional gene sets. However, the low reliability of the data and large sampling variability precludes making such estimate reliable. Thus, we next considered analyses that can provide estimates for the scope of post-transcriptional regulation even when the reliability of the data is low.

\subsection*{Coordinated post-transcriptional regulation of functional gene sets}
The low reliability of estimates across datasets limits the reliability of estimates of transcriptional and post-transcriptional regulation for individual proteins, \fg{reliability}. Thus, we focused on estimating the post-transcriptional regulation for sets of functionally related genes as defined by the gene ontology (GO) \citep{GO}. By considering such gene sets, we may be able to average out some of  the measurement noise and see regulatory trends shared by functionally related genes. Indeed, some of the noise contributing to the across-tissues variability of a gene is likely independent from the function of the gene; see Methods. Conversely, genes with similar functions are likely to be regulated similarly and thus have similar tissue-type-specific PTR ratios. Thus, we explored whether the across-tissues variability of the PTR ratios of functionally related genes reflects such tissue-type-specific and biological-function-specific post-transcriptional regulation.

Since this analysis aims to quantify across-tissues variability, we define the ``relative protein to mRNA ratio'' (rPTR) of a gene in a given tissue to be the PTR ratio in that tissue divided by the median PTR ratio of the gene across the other 11 tissues.  We evaluated the significance of rPTR variability for a gene-set in each tissue-type by comparing the corresponding gene-set rPTR distribution to the rPTR distribution for those same genes pooled across the other tissues (\fg{rPTR}); we use the KS-test to quantify the statistical significance of differences in the rPTR distributions; see Methods. The results indicate that the genes from many GO terms  have substantially higher rPTR in some tissues than in others. For example the ribosomal proteins of the small subunit (40S) have high rPTR in kidney but low rPTR in stomach (\fg{rPTR}a-c).

\blue{While the strong functional enrichment of rPTR suggests functionally concerted post-transcriptional regulation, it can also reflect systematic dataset-specific measurement artifacts. To investigate this possibility, we obtained two estimates for rPTR from independent datasets: Estimate 1 is based on data from \citet{wilhelm14}  and \citet{fagerberg2014analysis}, and Estimate 2 is based on data from \citet{kim2014draft} and \citet{ENCODE2012landscape}. These two estimates are reproducible (e.g., $\rho = 0.7-0.8$) for most tissues but less for others (e.g., $\rho = 0.14$), as shown by the scatter plots between the median rPTR for GO terms in \fg{rPTR}d;  \figs{3} shows the reproducibility for all tissues. The correlations between the two rPTR estimates remain statistically significant albeit weaker  (i.e., $\rho = 0.1-0.4$) when computed with all GO terms (not only those showing significant enrichment) as shown in Table \ref{rPTR_rGO_Table}, as well as when computed between the rPTRs for all genes,  Table \ref{rPTR_rGenes_Table}.}

\subsection*{Consensus protein levels }
Given the low reliability of protein estimates across studies \fg{reliability}, we sought to increase it by deriving consensus estimates. Indeed, by appropriately combining data from both protein studies, we can average out some of the noise thus improving the reliability of the consensus estimates; see Methods. 
As expected for protein estimates with increased reliability, the consensus protein levels correlate better to  mRNA levels than the corresponding protein levels estimated from a either dataset alone,  \fg{consensus}a,b.  We further validate our consensus estimates against 124 protein/tissue measurements from a targeted MS study by \citet{edfors2016gene}.  We computed the mean squared errors (MSE) between the protein levels estimated from the targeted study and the other three datasets using only protein/tissue measurements quantified in all datasets, facilitating fair comparison (\fg{consensus}c).  
The MSE are lower for the consensus dataset than for either \citep{wilhelm14} or \citep{kim2014draft} and are consistent with a 10\% error reduction relative to the \citet{kim2014draft}  dataset.  In addition to increased reliability, the consensus dataset increased coverage, providing a more comprehensive quantification of protein levels across human tissues than either draft of the human proteome taken alone (Table 1).

\section*{Discussion}  
Highly abundant proteins have highly abundant mRNAs. This dependence is consistently observed \citep{jovanovic2015dynamic, Csardi2015, gygi1999correlation, smits2014global, schwanhausser2011global} and dominates the explained variance in the estimates of absolute protein levels (\fg{figure1} and \figs{1}).  This underscores the role of transcription for setting the full dynamic range of protein levels.
In stark contrast, differences in the proteomes of distinct human tissues are poorly explained by transcriptional regulation, \fg{figure1}. This is due to measurement noise (\fg{reliability}) but also to post-transcriptional regulation. Indeed, large and \blue{partially} reproducible rPTR ratios suggest that the mechanisms shaping tissue-specific proteomes involve post-transcriptional regulation,  \fg{rPTR}. This result underscores the role of translational regulation and of protein degradation for mediating physiological functions within the range of protein levels consistent with life.

As with all analysis of empirical data, the results depend on the quality of the data and the estimates of their reliability. This dependence on data quality is particularly strong given that some conclusions rest on the failure of across-tissues mRNA variability to predict across-tissues protein variability. Such inference based on unaccounted for variability is substantially weaker than measuring directly and accounting for all sources of variability. 
The low across study reliability suggest that the signal is strongly contaminated by noise, especially systematic biases in sample collection and handling, and thus the data cannot accurately quantify the contributions of different regulatory mechanisms, \fg{reliability}.  Another limitation of the data is that isoforms of mRNAs and proteins are merged together, i.e., using razor proteins. This latter limitation is common to all approaches quantifying proteins and mRNAs from peptides/short-sequence reads. It stems from the limitation of existing approaches to infer and distinctly quantify isoforms and proteoforms.

The strong enrichment of rPTR ratios within gene sets (\fg{rPTR}) demonstrates a functionally concerted regulation at the post-transcriptional level. Some of the rPTR trends can account for fundamental physiological differences between tissue types. For example, the kidney is the most metabolically active (energy consuming) tissue among the 12 profiled tissues \citep{hall2010guyton} and it has very high rPTR for many gene sets involved in energy production (\fg{rPTR}a). In this case, post-transcriptional regulation likely plays a functional role in meeting the high energy demands of kidneys. Quantifying and understanding mRNA and protein covariation in single cells is an important frontier of this analysis \citep{ budnik2017SCoPE}.

The rPTR patterns and the across tissue correlations in \figs{1} indicate that the relative contributions of transcriptional and post-transcriptional regulation can vary substantially depending on the tissues compared. Thus, the level of gene regulation depends strongly on the context. For example transcriptional regulation is contributing significantly to the dynamical responses of dendritic cells \citep{jovanovic2015dynamic} and to the differences between kidney and prostate gland (\figs{1}h) but less to the differences between kidney and liver (\figs{1}g). All data, across all profiled  tissues, suggest that post-transcriptional regulation contributes substantially to the across-tissues variability of protein levels. The degree of this contribution depends on the context.

Indeed, if we only increase the levels for a set of mRNAs without any other changes, the corresponding protein levels must increase proportionally as demonstrated by gene inductions \citep{mcisaac2011fast}. However, the differences across cell-types are not confined only to different mRNA levels. Rather, these differences include different RNA-binding proteins, alternative untranslated regions (UTRs) with known regulatory roles in protein synthesis, specialized ribosomes \citep{mauro2002ribosome_filter, mauro2016translation, Slavov_ribo,preiss_2015}, and different protein degradation rates \citep{gebauer2004molecular, rojas2012alternative, castello2012insights, arribere2013roles, katz2014musashi}. The more substantial these differences, the bigger the potential for post-transcriptional regulation. Thus cell-type differentiation and commitment may result in much more post-transcriptional regulation than observed during perturbations preserving the cellular identity. 
Consistent with this possibility, tissue-type specific proteomes may be shaped by substantial post-transcriptional regulation; in contrast, cell stimulation that preserves the cell-type,   may elicit a strong transcriptional remodeling but weaker post-transcriptional remodeling.\\

\section*{Methods}

\subsection*{Data sources}
\blue{We used RNA estimates based on RNA-seq from \citet{fagerberg2014analysis, ENCODE2012landscape} and protein estimates based on shotgun mass-spectrometry from  \citet{wilhelm14, kim2014draft}. These large scale datasets contained $N = 6104$ genes measured in each of twelve different human tissues: adrenal gland, esophagus, kidney, ovary, pancreas, prostate, salivary gland, spleen, stomach, testis, thyroid gland, and uterus.  For these genes, about $8\%$ of the mRNA measurements and about $40\%$ of the protein measurements are missing. The mRNA datasets contain measurements from multiple subjects/people and the subjects were split into two subsets in estimating the within study reliability in \fg{reliability}a. We also used a small scale targeted dataset from \citet{edfors2016gene} containing data for 33 proteins measured across 5 tissues. The datasets were collected by different groups and measurements derived from different subjects. }

\subsection*{Searching raw MS data}
\blue{Raw data from \citet{kim2014draft, edfors2016gene} were searched by MaxQuant\citep{cox2008maxquant} 1.5.7.0 against a protein sequence database including all entries from a Human UniProt database from 2015 and known contaminants such as human keratins and common laboratory contaminants. MaxQuant searches were performed with trypsin specificity allowing up to two missed cleavages, with fixed Carbamidomethyl acetylation on cysteines, and with variable modifications allowing methionine oxidation and acetylation on Protein N-termminus. 
All razor peptides were used for quantifying the proteins to which they were assigned by MaxQuant. False discovery rate (FDR) was set to 1\% at both the protein and the peptide levels. }

\subsection*{Scaling mRNA levels}
First, denote $m_{it}$ the log mRNA levels for gene $i$ in tissue $t$.  Similarly, let $p_{it}$ denote the corresponding log protein levels.  First, we normalize the columns of the data, for both protein and mRNA, to different amounts of total protein per sample.  Any multiplicative factors on the raw scale correspond to additive constants on the log scale.   Consequently, we normalize data from each tissue-type by minimizing the absolute differences between data from the tissue and the first tissue (arbitrarily chosen as a baseline).  That is, for all $t>1$, we define
$$ p_{it}^n = (p_{it}^u - \hat{\mu}_t)$$
with
$$ \hat{\mu}_t = \underset{\mu}{\text{argmin}} \sum_i |p_{i1}^u -(p_{it}^u - \mu)|$$
\noindent Where $p_{it}^n$ and $p_{it}^u$ represent the normalized and non-normalized protein measurements respectively.  For each $t$, the value of $\mu_t$ which minimizes the absolute difference is
$$\hat{\mu}_t = \underset{u}{\text{median}} (p_{i1}-p_{it}^u) $$
We use the same normalization for mRNA. This normalization, which corresponds to a location shift of the log abundances for each tissue, corrects for any multiplicative differences in the raw (unlogged) mRNA or protein. We normalize these measurements by aligning the medians rather than the means, as the median is more robust to outliers.

After normalization, we define $r_{it} = p_{it}- m_{it}$ as the log PTR ratio of gene $i$ in condition $t$.  If the post-transcriptional regulation for the $i^{th}$ gene were not tissue-specific, then the $i^{th}$ PTR ratio would be independent of tissue-type and can be estimated as $$ \hat{T}_i = \underset{t}{\text{median}}(p_{it}- m_{it})$$

\noindent In such a situation the log ``scaled mRNA'' (or mean protein level) can be defined as

$$\overline{p}_{it} = m_{it}+T_i $$

\noindent On the raw scale this amounts to scaling each mRNA by its median PTR ratio and represents and estimate of the mean protein level.  The residual difference between the log mean protein level and the measured log protein level, which we call the log rPTR ratio

$$ r_{it} = p_{it}-\overline{p}_{it}$$

\noindent consists of both tissue-specific post-transcriptional regulation and measurement noise.

\subsection*{Across-tissues correlations}
For each gene, $i$, we compute the correlation between mRNA and protein across tissues.  Unlike the between gene correlations which are consistently large after scaling for each tissue (Figure \ref{figure1}a), across-tissues correlations are highly variable between genes.  Although this could be in part because true mRNA/protein correlations vary significantly between genes, a huge amount of the heterogeneity can be explained by sampling variability.  There are only 10 and 12 tissues in common across datasets (depending on which datasets are used) and for many genes the abundances are missing, which means that the empirical estimates of across tissue correlation for each gene are very noisy.  To find a representative estimate of the across-tissues correlation we can take the median over all genes.  As an alternative, if the correlation was roughly constant between genes, we could pool information to yield a representative estimate of this across-tissues correlation. For a gene $i$, we compute the Fisher transformation of the within-gene correlation.  This Fisher transformation, $z_i = \text{arctanh}(\hat{r}_i)$ is approximately normally distributed:
 $$z_i \sim N\left(\frac{1}{2}log(\frac{1+\rho}{1-\rho}),\frac{1}{\sqrt{N_i-3}}\right)$$
where $N_i$ are the number of observed mRNA-protein pairs for gene $i$ (at most 11) and $\rho$ corresponds to the population correlation.  We estimate the maximum likelihood estimate of the Fisher transformed population correlation by weighting each observation by its variance:
\begin{eqnarray*}
\omega_i &=& \frac{1}{n_i-3}\\
W_i &=& \frac{\omega_i}{\sum_j \omega_j}\\
\hat{z}_{pop}  &=& \sum W_iz_i\\
\end{eqnarray*}
We then transform this estimate back to the correlation scale
\begin{eqnarray*}
\hat{\rho} &=&\frac{e^{2\hat{z}_{pop}}-1}{e^{2\hat{z}_{pop}}+1}
\end{eqnarray*}
Depending on the data sets used, with this method we estimate the population across-tissues mRNA/protein correlation to be between 0.21 (\citep{wilhelm14}) and 0.29 (\citep{kim2014draft}).  This correlation \emph{cannot} be used as direct evidence for the relationship between mRNA and protein levels since both mRNA and protein datasets are unreliable due to measurement noise.  This measurement noise attenuates the true correlation.  Below we address this by directly estimating data reliability and correcting for noise.

\subsection*{Noise correction}
Measurement noise attenuates estimates of correlations between mRNA and protein level \citep{Franks2014}. A simple way to quantify this attenuation of correlation due to measurement error is via Spearman's correction.  Spearman's correction is based on the fact that the variance of the measured data can be decomposed into the sum of variance of the noise and the signal. If the noise and the signal are independent, this decomposition and the Spearman's correction are exact \citep{Csardi2015}. 

Note that it is simple to show that the empirical variance is the sum of the variance of the signal and the variance of the noise: 		
\begin{itemize}
 	 \item $e_i$ - Expectation at the $i^{th}$ data point; $\tilde e_i$ = $e_i - \langle e \rangle$
 	 \item $\zeta_i$ - Noise at the $i^{th}$ data point; $\langle \zeta \rangle = 0$
 	 \item $x_i$ - Observation at the $i^{th}$ data point; $\tilde x_i$ = $x_i - \langle x \rangle$, $x_i = e_i + \zeta_i$;
 \end{itemize}
 \begin{align*}
  	& \sigma_x^2  = \frac{1}{n}\sum_i \tilde x_i^2 = \frac{1}{n} \sum_i \left( \tilde e_i + \zeta_i \right)^2 =  \\
  	&						  = 
  	\underbrace{ \frac{1}{n} \sum_i \tilde e_i^2 }_{\sigma_e^2} +
  	\underbrace{ \frac{1}{n} \sum_i  \zeta_i^2 }_{\sigma_{\zeta}^2} +
  	\underbrace{ \frac{2}{n}\sum_i \tilde e_i \zeta_i }_{\approx 0}
 \end{align*}
Spearman's correction is based on estimates of the ``reliability'' of the measurements, which is defined as the fraction of total measured variance due to signal rather than to noise:
\begin{eqnarray}
\label{eqn:rel}
 \text{Reliability}  &=& \frac{\sigma^2_{signal}}{\sigma^2_{total}}\\
&=& 1 - \frac{\sigma^2_{err}}{\sigma^2_{err}+\sigma^2_{signal}}
\end{eqnarray}
If $X$ and $Y$ are noisy measurements of two quantities, we can compute the noise corrected correlation between them as
\begin{eqnarray}
\label{eqn:correction}
\frac{Cor(X,Y)}{\sqrt{Rel(X)Rel(Y)}}\\
\end{eqnarray}
In practice, reliabilities are not known but we can often estimate them.  In this application, for both mRNA and protein we need measurements in which all steps, from sample collection to level estimation, are repeated independently.  In order to estimate the mRNA reliabilities we use independent measurements from \citet{fagerberg2014analysis} and \citet{ENCODE2012landscape}.  For estimating protein reliabilities we use measurements from \citet{wilhelm14} and \citet{kim2014draft}.  across-tissues reliabilities are computed per gene whereas within-tissue reliabilities are computed per tissue across genes.  If two independent measurements have the same reliability, it can be estimated by computing the correlation between the two measurements \citep{spearman04, zimmerman97, Csardi2015}.  We estimated the approximate across-tissues protein reliability to be 0.21 and the across-tissues mRNA reliability to be 0.77.  Given the estimated across-tissues mRNA/protein correlation of 0.29 (calculated using data from \citet{kim2014draft} and \citet{fagerberg2014analysis}) we estimated the noise-corrected fraction of across-tissues protein variance explained by mRNA to be approximately 50\%, Figure \ref{reliability}. Note that if both mRNA or both protein datasets share biases, then the estimated reliabilities will be too small, thus deflating the inferred fraction of protein variance explained by mRNA.  Moreover, because the reliabilities are low, sampling variability is large, missing data is prevalent, and mRNA/protein correlation likely vary by gene there is uncertainty about this estimate.

\subsection*{Creating a consensus protein dataset}
We use the two independent protein datasets to create a single consensus data set which is of arguably higher reliability than either dataset individually.   To create this dataset, we take a weighted average of the two protein abundance datasets, by tissue.  We compute the weights based on measurement reliabilities for each tissue in each of the two datasets.  

Assume we have two random variables, $\overset{\sim}{X}_1$ and $\overset{\sim}{X}_2$, corresponding to measurements on the same quantity (e.g. two independent protein measurements) with $\overset{\sim}{X}_i = X + \epsilon_i$ where  $X \sim N(0, \sigma_X^2)$ is the signal which is independent of $\epsilon_i \sim N(0, \sigma^2_{\epsilon_i})$, the measurement error for sample $i$.  We have a third random variable corresponding to a different quantity (e.g. an mRNA measurement), $\overset{\sim}{Y}$ that is typically positively correlated with $\overset{\sim}{X}_1$ and $\overset{\sim}{X}_2$ with the same covariance $\sigma^2_{XY}$.  To create the consensus data set we first compute the reliability of $\overset{\sim}{X}_i$ $Rel(\overset{\sim}{X}_i) = \frac{\sigma_X^2}{\sigma^2_{\overset{\sim}{X}_i}} = \frac{\sigma^2_X}{\sigma^2_X + \sigma^2_{\epsilon_i}}$ for both datasets.

Note that
\begin{eqnarray*}
\text{Cor}(\overset{\sim}{X}_1, \overset{\sim}{X}_2) &=& \frac{\sigma_X^2}{\sigma_{\overset{\sim}{X}_1}\sigma_{\overset{\sim}{X}_2}}\\
\text{Cor}(\overset{\sim}{X}_i, Y) &=& \frac{\sigma^2_{XY}}{\sigma_{\overset{\sim}{X}_i}\sigma_{Y}}\\
\end{eqnarray*}

Thus, 
\begin{eqnarray*}
\text{Cor}(\overset{\sim}{X}_1, \overset{\sim}{X}_2)\frac{\text{Cor}(\overset{\sim}{X}_1, \overset{\sim}{Y})}{\text{Cor}(\overset{\sim}{X}_2, \overset{\sim}{Y})} &=& \frac{\sigma_X^2}{\sigma^2_{\overset{\sim}{X}_1}}\\
&=&\frac{\sigma_X^2}{\sigma^2_X + \sigma^2_{\epsilon_1}}\\
&=& Rel(\overset{\sim}{X}_1)
\end{eqnarray*}

\noindent Similarly, $\text{Cor}(\overset{\sim}{X}_1, \overset{\sim}{X}_2)\frac{\text{Cor}(\overset{\sim}{X}_2, \overset{\sim}{Y})}{\text{Cor}(\overset{\sim}{X}_1, \overset{\sim}{Y})} = \text{Rel}(\overset{\sim}{X}_2)$.  We use these facts and compute the empirical correlations between datasets to independently estimate the across gene reliabilities for each tissue from each dataset.  We then Fisher weight the protein abundances based on their reliabilities.  That is, for each tissue $t$, the consensus dataset, $X^C_{t}$ is
\begin{eqnarray*}
X^C_{t} &=& w\overset{\sim}{X}^t_{t} + (1-w)\overset{\sim}{X}^2_{t}\\
w &=& \frac{Rel(\overset{\sim}{X}^1_{t})}{Rel(\overset{\sim}{X}^1_{t}) + Rel(\overset{\sim}{X}^2_{t})}
\end{eqnarray*}
When the reliability of $\overset{\sim}{X}^1_{t}$ and $\overset{\sim}{X}^2_{t}$ are close, each dataset is weighted equally.  When one reliability dominates the other, that dataset contributes more to the aggregated dataset.  We found that the full consensus data set has a higher median per gene correlation with mRNA than either of the protein datasets individually (0.34) and agreed more closely with validation data from \citep{edfors2016gene} (Table \ref{mseTab}).

\subsection*{Functional gene set analysis}

To identify tissue-specific rPTR for functional sets of genes, we analyzed the distributions of rPTR ratios within functional gene-sets using the same methodology as \citet{Slavov_eth_grr}.  We restrict our attention to functional groups in the GO ontology \citep{GO} for which at least 10 genes were quantified by \citet{wilhelm14}.  Let $k$ index one of these approximately 1600 functional gene sets.  First, for every gene in every tissue we estimate the relative PTR (rPTR) or equivalently, the difference between log mean protein level and measured protein level:
$$\hat{r}_{it}= p_{it} - \underset{t'\neq t}{\text{median}}(p_{it'}- m_{it'})$$

To exclude the possibility that $\hat{r}_{it}=0$ exactly, we require that $t' \neq t$.  When the estimated rPTR is larger than zero, the measured protein level in tissue $t$ is larger than the estimated mean protein level.   Likewise, when this quantity is smaller than zero, the measured protein is smaller than expected.  Measured deviations from the mean protein level are due to both measurement noise and tissue specific PTR.  To eliminate the possibility that all of the variability in the rPTR ratios is due to measurement error we conduct a full gene set analysis.

For each of the gene sets we compute a vector of these estimated log ratios so that a gene set is comprised of

$$\mathcal{G}_{kt} = \{\hat{r}_{i_1j},...,\hat{r}_{i_{n_k}t}\}$$

where $i_1$ to $i_{n_k}$ index the genes in set $k$ and $t$ indexes the tissue type. 

Let $KS(\mathcal{G}_1,\mathcal{G}_2)$ be the function that returns the p-value of the Kolmogorov-Smirnov test on the distribution in sets $\mathcal{G}_1$ and $\mathcal{G}_2$. The KS-test is a test for a difference in distribution between two samples.  Using this test, we identify gene sets that show systematic differences in PTR ratio in a particular tissue ($t$) relative to all other tissues.  

Specifically, the p-value associated with  gene set $k$ in condition $j$ is 
$$\rho_{kt} = \text{KS}(\mathcal{G}_{kt},\underset{t' \neq t}{\cup} \mathcal{G}_{kt'})$$

To correct for testing multiple hypotheses, we computed the false discovery rate (FDR) for all gene sets in tissue $t$ \citep{storey2003}.  In Figure \ref{rPTR}a-c, we present only the functional groups with FDR less than 1\% and report their associated p-values. Note that the test statistics for each gene set are positively correlated since the gene sets are not disjoint, but \citet{benjamini2001control} prove that the Benjamini-Hochberg procedure applied to positively correlated test statistics is conservative.  Thus, the significance of of certain functional groups suggests that not all of the variability in rPTR is due to measurement noise.  We also calculated rPTR using two pairs of measurements: one set of rPTR estimates was calculated using protein data from \citet{wilhelm14} and mRNA from \citet{fagerberg2014analysis} and the other was calculated using data from \citet{kim2014draft} and \citet{ENCODE2012landscape}.  rPTR of the significant sets was largely reproducible across estimates from independent datasets (Figure \ref{rPTR}d) and less reproducible across genes (Table \ref{rPTR_rGenes_Table}).   Note that when computing the per tissue reliabilities for the construction of the consensus data set, we found that the reliabilities of the lung and pancreas datasets from \citet{wilhelm14} were much less reliable than the data from \citet{kim2014draft}.  This could explain why the independent estimates of the rPTR ratios for these tissues were less reproducible.

\noindent{\bf Acknowledgments}
We thank M.~Jovanovic, H.~Specht, E.~Wallace, J.~Schmiedel, and D.~A.~Drummond for discussions and constructive comments. 


\noindent {\bf Supplemental website: }
 \href{https://alum.mit.edu/www/nslavov/2016_PTR/}{https://web.northeastern.edu/slavovlab/2016\_PTR/}  

\noindent {\bf The code can be found at: }
 \href{https://github.com/afranks86/tissue-ptr}{https://github.com/afranks86/tissue-ptr}  
   

\section*{\Huge Tables}
\begin{table}[ht]
\centering
\begin{tabular}{lcccccc}
  \hline
  \rowcolor{lightGray}
 & \multicolumn{2}{c}{Wilhelm et al.}
 & \multicolumn{2}{c}{Kim et al.}
 & \multicolumn{2}{c}{Consensus} \\
 \rowcolor{lightGray}
 Tissue type & Proteins & Peptides & Proteins & Kim pep & Proteins & Peptides \\
  \rowcolor{white}
Adrenal Gland & 3,947 & 43,661 & 3,610 & 40,619 & 4,436 & 62,719 \\ 
  Colon & 3,745 & 34,925 & 4,004 & 32,786 & 4,517 & 52,872 \\ 
  Esophagus & 3,989 & 39,677 & 1,682 & 15,298 & 4,086 & 45,881 \\ 
  Kidney & 3,045 & 34,740 & 2,814 & 26,499 & 3,526 & 45,865 \\ 
  Liver & 3,669 & 34,844 & 2,307 & 22,992 & 3,971 & 45,168 \\ 
  Lung & 2,852 & 31,293 & 2,935 & 29,686 & 3,704 & 47,617 \\ 
  Ovary & 3,681 & 41,289 & 3,568 & 53,956 & 4,419 & 72,337 \\ 
  Pancreas & 3,048 & 30,463 & 2,326 & 19,974 & 3,423 & 38,807 \\ 
  Prostate & 3,467 & 38,562 & 3,118 & 33,765 & 3,932 & 51,891 \\ 
  Testis & 3,950 & 51,369 & 3,505 & 42,865 & 4,478 & 70,952 \\ 
  Spleen & 3,244 & 39,465 & 0 & 0 & 3,237 & 39,465 \\ 
  Stomach & 3,957 & 44,609 & 0 & 0 & 3,949 & 44,609 \\ 
  Heart & 3,957 & 0 & 2,227 & 27,205 & 2,227 & 27,205 \\ 
  All & 5,719 & 113,069 & 5,326 & 105,158 & 5,786 & 150,830 \\ 
   \hline
\end{tabular}
\caption{ \covarage_Table } 
\label{mseTab}
\end{table}
 \newpage

\section*{\Huge Figures}

\setcounter{figure}{0}  

\begin{figure}[h!]
\includegraphics[scale=0.88]{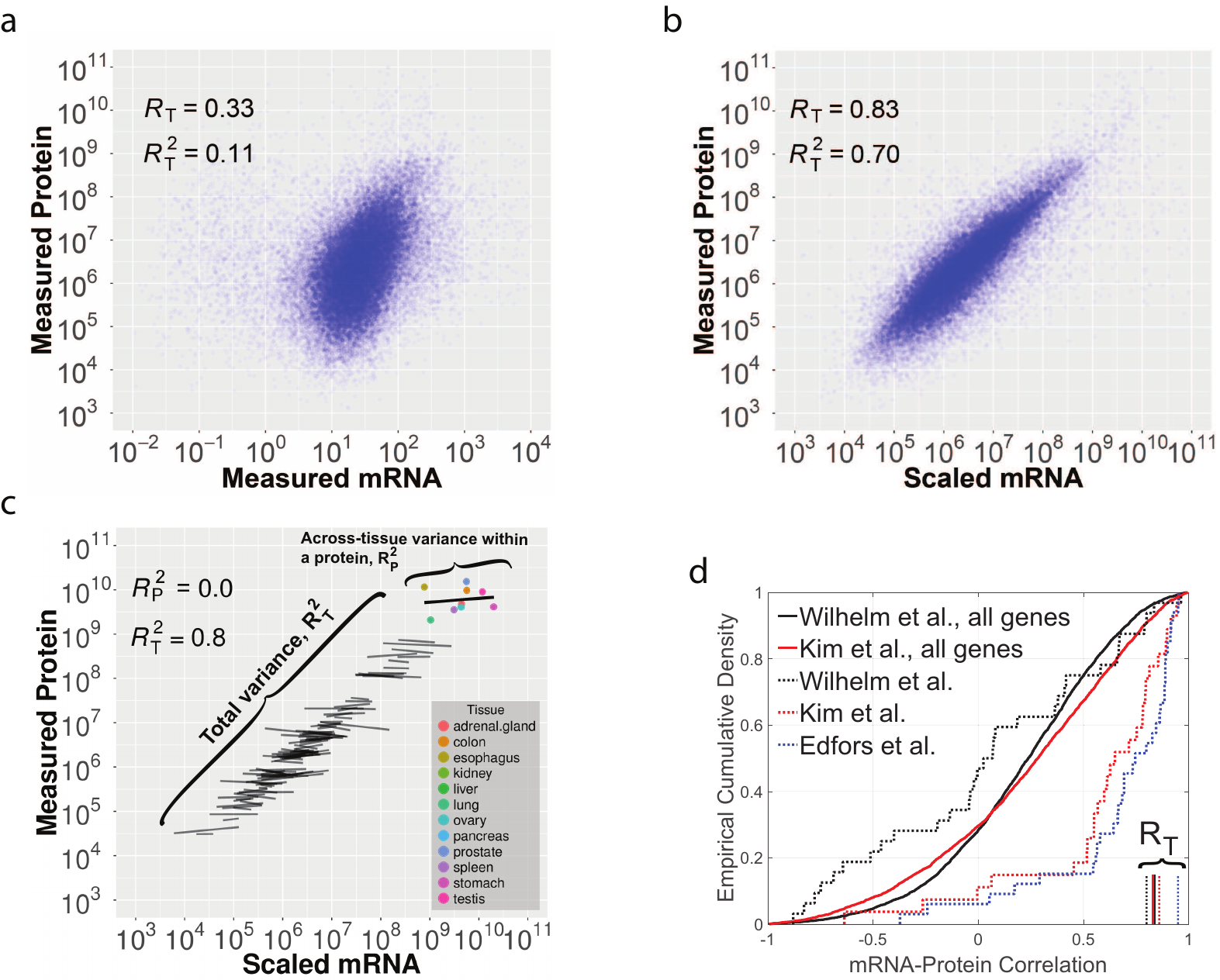}
\caption{ \simParadox } 
\label{figure1}
\end{figure}
\newpage

\begin{figure}[h!]
\includegraphics[scale=0.88]{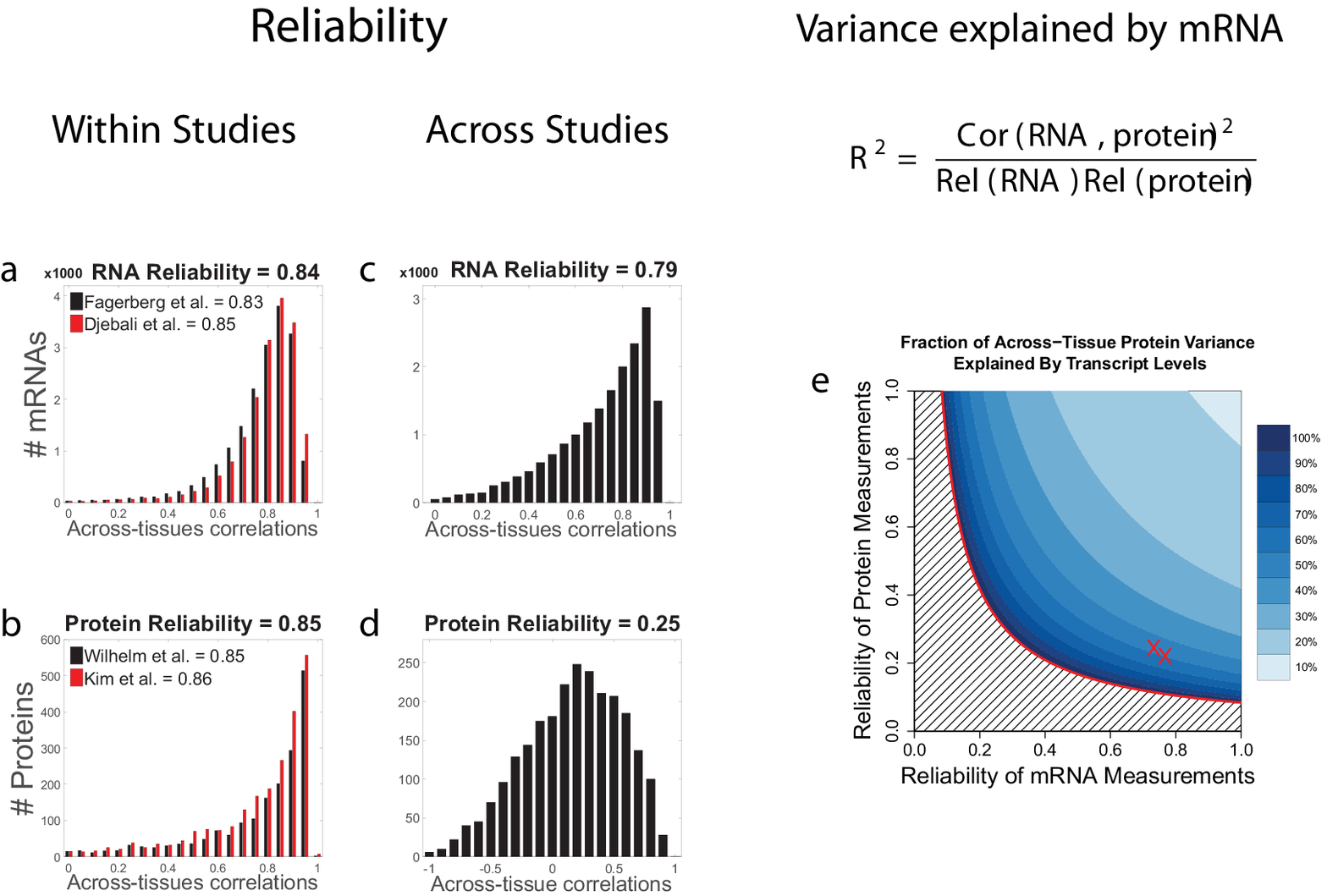}
\caption{ \Reliability }  
\label{reliability}
\end{figure}
\newpage

\begin{figure}[h!]
\includegraphics[scale=0.88]{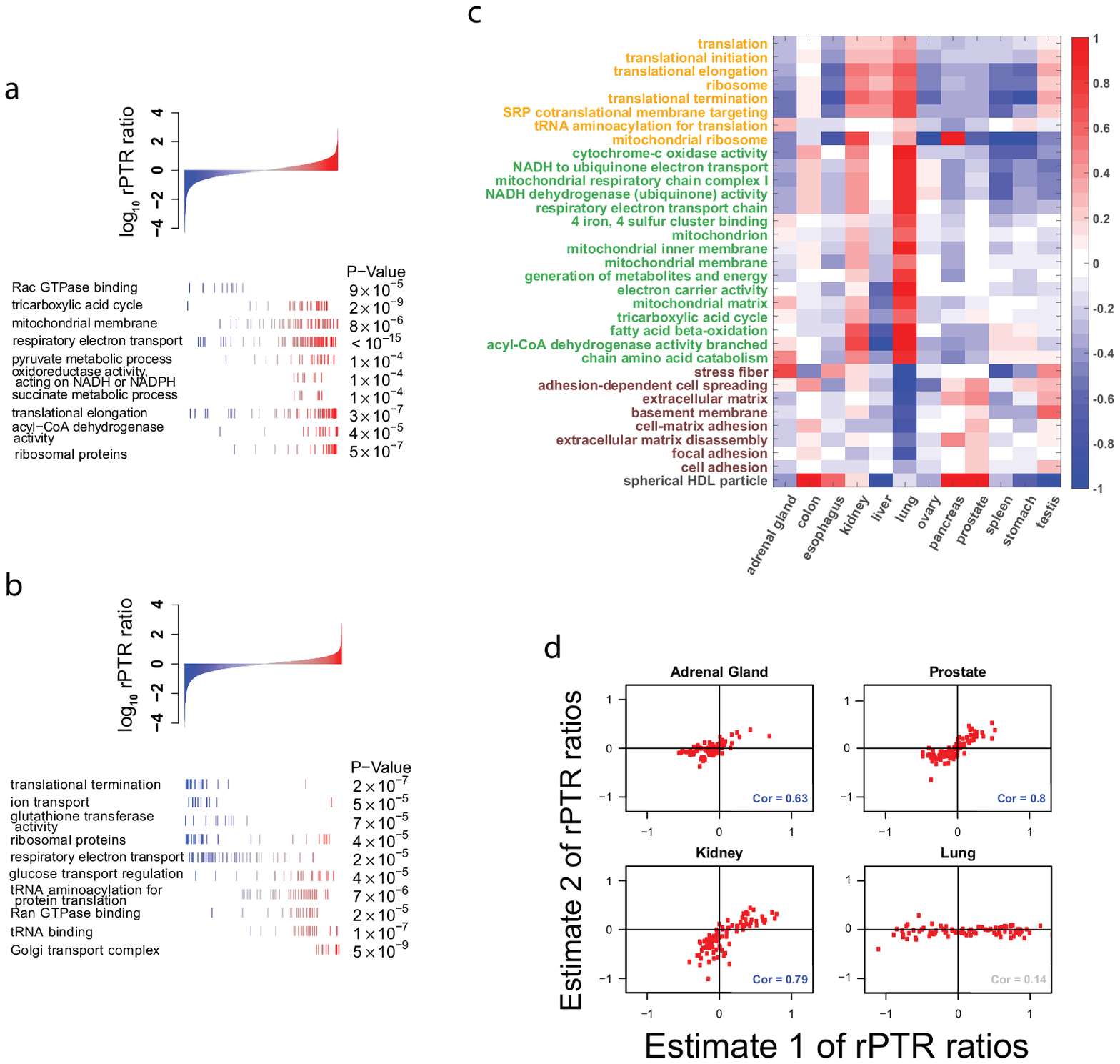}
\caption{ \rPTR }   
\label{rPTR}
\end{figure}
\newpage

\begin{figure}[h!]
\includegraphics[scale=0.88]{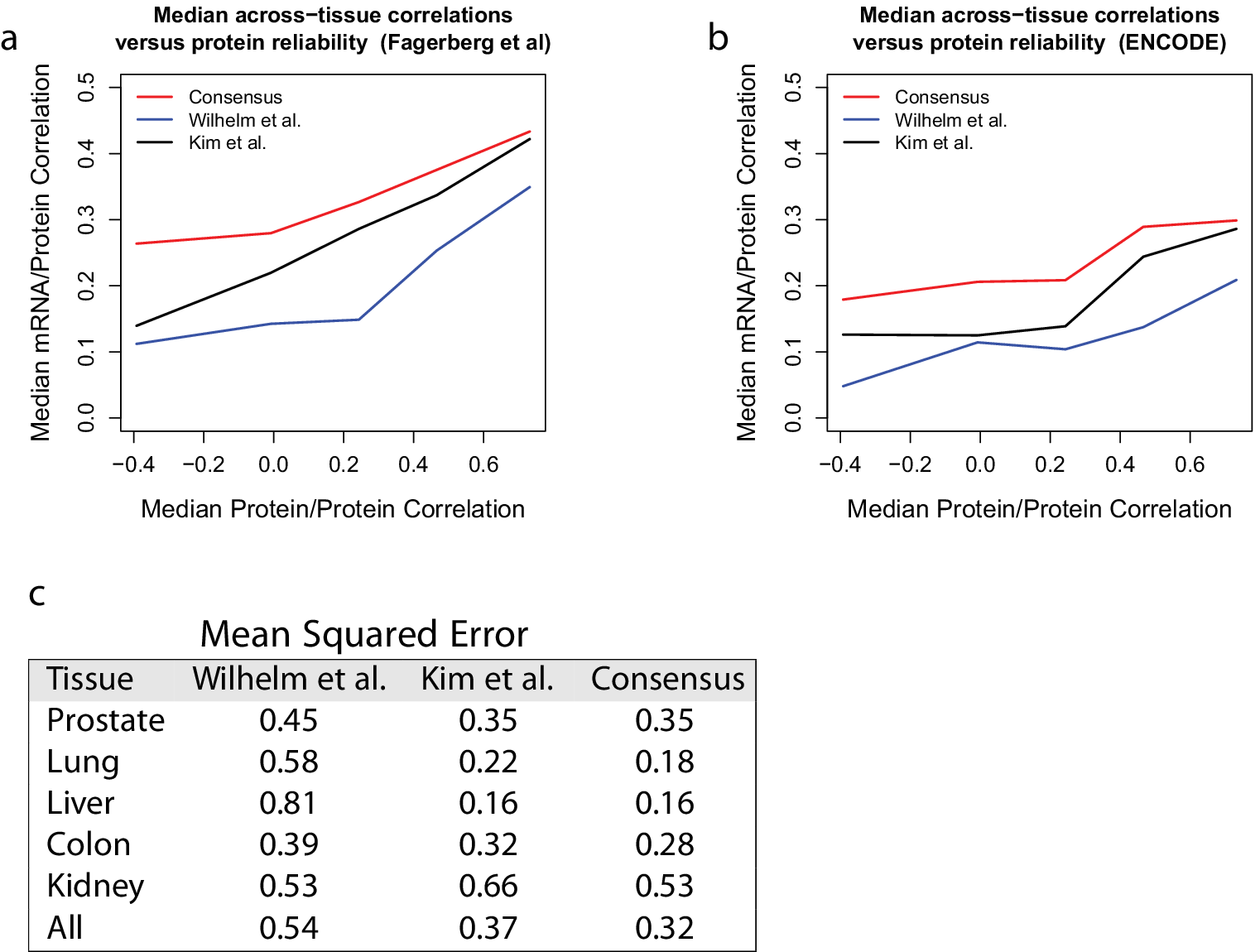}
\caption{ \consensus }   
\label{consensus}
\end{figure}
\clearpage

\end{spacing} 
\setcounter{table}{0}		\renewcommand{\thetable }{ S\arabic{table}}

\section*{\Huge Supplemental Tables and Figures }

\begin{table}[ht]
\caption{ \rPTRrGOTable }   
\label{rPTR_rGO_Table}
\end{table}

\begin{table}[ht]
\caption{ \rPTRrGenesTable }   
\label{rPTR_rGenes_Table}
\end{table}

\noindent
{\bf Dataset S1. Consensus dataset of protein levels across human tissues}
    A zip-archived comma-delimited text file with consensus estimates of protein levels across 13 human tissues: adrenal gland,	colon,	esophagus,	kidney,	liver,	lung,	ovary,	pancreas,	prostate,	testis,	spleen,	stomach, and heart.\\

\noindent
{\bf Dataset S2. Peptide levels across human tissues}
    A zip-archived comma-delimited text file with estimates of peptide levels across 13 human tissues: adrenal gland,	colon,	esophagus,	kidney,	liver,	lung,	ovary,	pancreas,	prostate, testis,	spleen,	stomach, and heart. This file contains all peptide levels (integrated precursors areas) estimated from the MaxQuant searches described in the Methods. 
 

\setcounter{figure}{0}      \renewcommand{\thefigure}{ S\arabic{figure}}

\begin{figure}[h!]
 \caption{ \Simpson_Supp }
\label{Simpson_Supp}
\end{figure}

\begin{figure}[h!]
    \caption{ \GOterm_Corrs }
\label{figure5}
\end{figure}

\begin{figure}[h!]
    \caption{ \rptrCorrs }
\label{figure6}
\end{figure}

\end{document}